\documentclass[lettersize,journal]{IEEEtran}

\usepackage{graphicx}
\usepackage{citesort}
\usepackage{amssymb}
\usepackage{amsfonts}
\usepackage{amsmath}
\usepackage{color}
\usepackage{fancybox}
\usepackage{textcomp}
\usepackage{multirow}
\usepackage{setspa ce}
\usepackage{psfrag}
\usepackage[ruled,vlined,linesnumbered]{algorithm2e}

\usepackage{mathrsfs}

\usepackage{booktabs}
\usepackage{threeparttable}

\def\BibTeX{{\rm B\kern-.05em{\sc i\kern-.025em b}\kern-.08em
    T\kern-.1667em\lower.7ex\hbox{E}\kern-.125emX}}
\usepackage{balance}

    \def\Complex{{\rm\rule[.23ex]{.03em}{1.1ex}\kern-.3em{C}}}

    \newcommand{\be}{\begin{equation}} \newcommand{\ee}{\end{equation}}
    \newcommand{\bea}{\begin{eqnarray}} \newcommand{\eea}{\end{eqnarray}}
    \newcommand{\benum}{\begin{enumerate}} \newcommand{\eenum}{\end{enumerate}}

    \newcommand{\qn}{{\bf n}}

    \newcommand{\qs}{{\bf s}}

    \newcommand{\qv}{{\bf v}}

    \newcommand{\qy}{{\bf y}}
    \newcommand{\qz}{{\bf z}}

    \newcommand{\qH}{{\bf H}}
    \newcommand{\qI}{{\bf I}}

    \newcommand{\qR}{{\bf R}}

    \newcommand{\qPhi}{{\boldsymbol \Phi}}

    \newcommand{\bbE}{{\mathbb E}}

    \newcommand{\bbC}{{\mathbb C}}

    \newcommand{\tr}{{\sf tr}}

\begin{document}

\title{Active RIS-Assisted MIMO-OFDM System: Analyses and Prototype Measurements}

\author{De-Ming Chian, \IEEEmembership{Graduate Student Member,~IEEE}, Feng-Ji~Chen, Yu-Chen~Chang,\\
~Chao-Kai~Wen, \IEEEmembership{Senior Member,~IEEE},~Chi-Hung~Wu,~Fu-Kang~Wang,~\IEEEmembership{Member,~IEEE}, Kai-Kit~Wong,~\IEEEmembership{Fellow,~IEEE},~and~Chan-Byoung~Chae,~\IEEEmembership{Fellow,~IEEE}

\thanks{D.-M.~Chian, F.-J.~Chen, Y.-C.~Chang, and C.-K.~Wen are with the Institute of Communications Engineering, National Sun Yat-sen University, Kaohsiung 804, Taiwan, Email: {\rm icefreeman123@gmail.com}, {\rm king19635@gmail.com}, {\rm william881106@gmail.com}, {\rm chaokai.wen@mail.nsysu.edu.tw}.}
\thanks{C.-H.~Wu and F.-K.~Wang are with the Department of Electrical Engineering, National Sun Yat-sen University, Kaohsiung 804, Taiwan, Email: {\rm yes121335@gmail.com}, {\rm fkw@mail.ee.nsysu.edu.tw}.}
\thanks{{K.-K.~Wong} is with Department of Electronic and Electrical Engineering, University College London, UK, Email: {\rm kai-kit.wong@ucl.ac.uk}. He is also affiliated with Yonsei Frontier Lab., Yonsei University, Korea.}
\thanks{{C.-B.~Chae} is with the School of Integrated Technology, Yonsei University, Seoul 03722, Korea, Email: {\rm cbchae@yonsei.ac.kr}.}
}

\markboth{IEEE Communications Letters,~Vol.~XX, No.~XX, XXX~2023}%
{Shell \MakeLowercase{\textit{et al.}}: Bare Demo of IEEEtran.cls for Journals}

\maketitle

\begin{abstract}

In this study, we develop an active reconfigurable intelligent surface (RIS)-assisted multiple-input multiple-output orthogonal frequency division multiplexing (MIMO-OFDM) prototype compliant with the 5G New Radio standard at 3.5~GHz.
The experimental results clearly indicate that active RIS plays a vital role in enhancing MIMO performance, surpassing passive RIS. Furthermore, when considering factors such as complexity, energy consumption, and performance, the comparative evaluation between passive RIS and active RIS reinforces the critical role of active RIS in MIMO systems. These findings underscore the practical significance of active RIS in improving MIMO gain in 5G scenarios.
\end{abstract}


\section*{I. Introduction}
\IEEEPARstart{F}{uture} wireless communication systems require advanced transceiver designs and support for dynamic wireless channel transformations. Reconfigurable Intelligent Surfaces (RIS) have emerged as a promising solution in achieving this goal. Extensive prototypes and experiments have been conducted in academia and industry, as evidenced by the works~\cite{Dai-20IAccess,Pei-21TCOM,Fara-22WCOM,Araghi-22IAccess,Sang-22WCOM,Ren-2022,Chian-2023}.
Despite their potential, it is known that the cascaded path loss of the RIS link can be approximated as the product of the path losses of the transmitter-RIS link and the RIS-receiver link, resulting in weaker multipath characteristics compared to the direct link. Traditional approaches involve equipping RIS with large reflecting elements, which introduce challenges in terms of channel estimation and real-time optimization of reflection coefficients \cite{Kang-arXiv23}.
To address these challenges, the concept of active RIS has been proposed, enabling simultaneous signal reflection and amplification. Unlike passive RIS systems, active RIS systems provide more effective compensation for severe cascaded path loss \cite{Long-21TWC,Kang-arXiv23,Zhang-23TCOM,Fu-Mul_ARIS/PRIS,Li-23COML}.

So far, most RIS-assisted wireless communication prototypes and measurements have focused on single-input single-output (SISO) scenarios using horn antennas. This configuration, primarily suitable for directional communications, does not align with the more omnidirectional nature of practical base station (BS) and smartphone form factors, especially for Sub-6 GHz communications.
There is a scarcity of information available regarding RIS-aided multiple-input multiple-output (MIMO) systems.
This study aims to address this gap by presenting the results of an experimental demonstration of an active RIS-assisted MIMO-OFDM prototype operating at 3.5 GHz under the 5G New Radio (NR) standard.
The key contributions of this study are as follows:
\begin{itemize}
\item This study marks the first experimental development of active RIS-assisted MIMO-OFDM within 5G NR, providing insights into RIS implementation in MIMO systems.
\item We introduce a unified platform that implements various RIS-assisted MIMO control algorithms, such as codebook-based, conditional sample mean (CSM) \cite{Ren-2022}, and blind greedy (BG) algorithms \cite{Chian-2023}. This platform facilitates real-world comparisons of different RIS-assisted MIMO methodologies.
\item We deliver a thorough evaluation comparing passive and active RIS using different elements and algorithms. This comparison elucidates the trade-offs in complexity, energy consumption, and performance, offering insights into the distinct features of these RIS technologies.
\end{itemize}
The experimental results highlight the greater potential of active RIS over passive RIS, particularly in the context of MIMO systems.

\begin{figure}
    \centering
    \resizebox{3.0in}{!}{%
    \includegraphics*{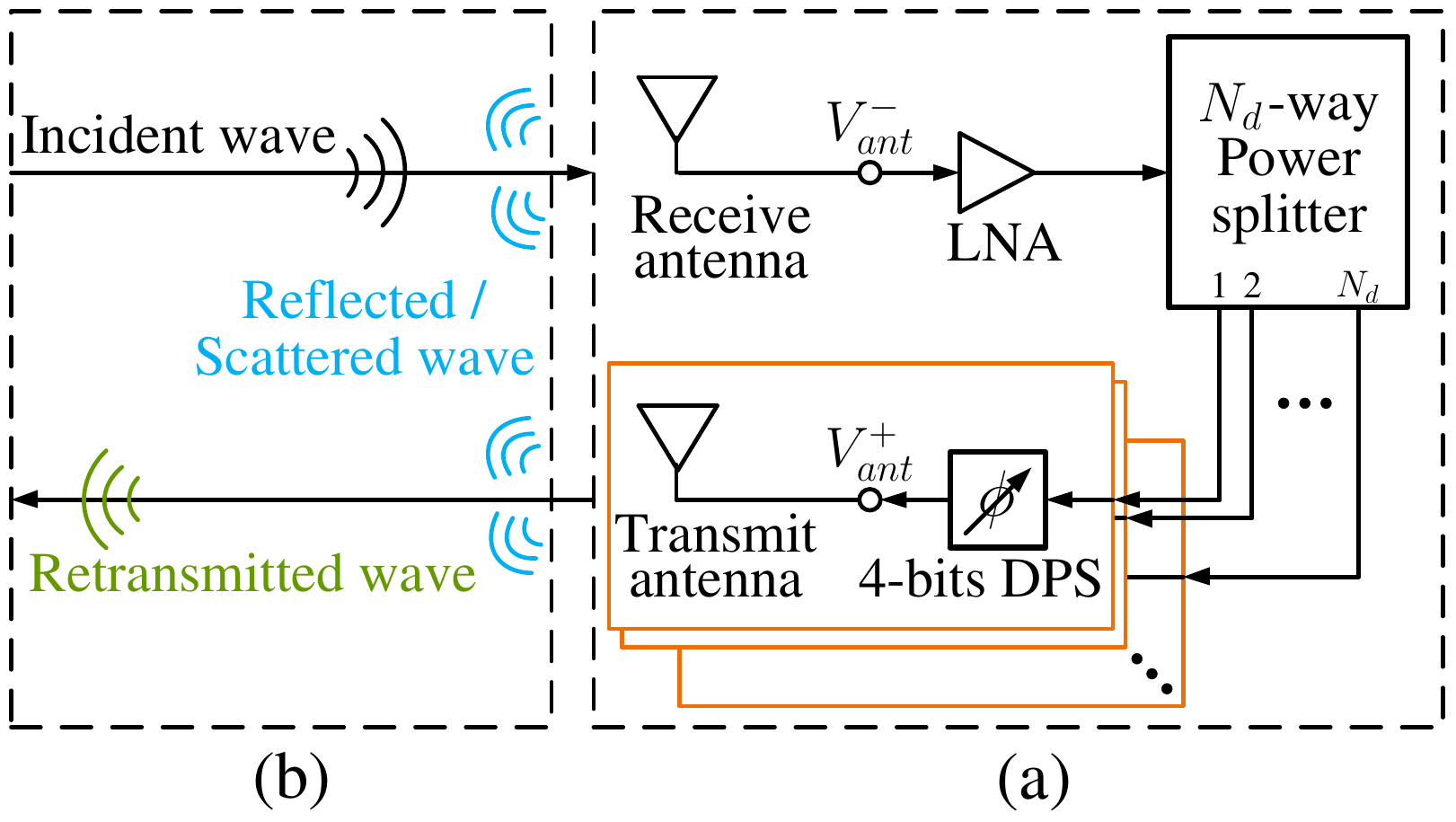} }%
    \caption{(a) Active RIS element and (b) its channel model.
    \label{fig:RISelement}}
    \vspace{-0.5cm}
\end{figure}

\section*{II. Design of Active RIS}

\subsection{Active RIS Element}
The structure of our ``active RIS element,'' as depicted in Fig. \ref{fig:RISelement}(a), consists of a receive antenna (rRIS) and $N_{\rm d}$ transmit antennas (tRIS). When an electromagnetic wave is incident on the rRIS, it passes through a low-noise amplifier (LNA) and an $N_{\rm d}$-way power splitter (PS). Each of the $N_{\rm d}$ separated paths is then individually adjusted by a digital phase shifter (DPS) and subsequently radiated back into the air through the $N_{\rm d}$ tRIS antennas. The rRIS and tRIS antennas are based on patch antenna technology, similar to the approach described in~\cite{Chian-2023}. To mitigate self-interference, we used separate antennas for tRIS and rRIS. The operating center frequency of our system is 3.5 GHz. The DPS, {\tt MAPS-010144}, is a 4-bit phase shifter with 5 \textmu W power consumption manufactured by M/A-COM Technology Solutions Inc., capable of providing phase shifts ranging from $0^\circ$ to $360^\circ$ in steps of $22.5^\circ$. The LNA, {\tt ZX60-83LN-S+}, with a gain of approximately 20 dB and 300 mW power consumption, and the PS, {\tt ZB8PD-622-S+}, with 8 output ports, are sourced from Mini-Circuits Inc.

By excluding the rRIS and tRIS elements in Fig. \ref{fig:RISelement}(a), the total transmission coefficient from the LNAs to the DPS can be expressed as 
\begin{equation} \label{eq:TC}
    \Phi = \frac{V_{\rm ant}^{+}}{V_{\rm ant}^{-}}
    = \Phi_{\rm lna} \cdot \Phi_{\rm ps} \cdot \Phi_{\rm dps},
\end{equation}
where $V_{\rm ant}^{+}$ and $V_{\rm ant}^{-}$ $\in \bbC$ denote the signal traveling towards and away from the antenna port, respectively, and $\Phi_{\rm lna}$, $\Phi_{\rm ps}$, $\Phi_{\rm dps}$ $\in \bbC$ are the transmission coefficients of the LNA, PS, and DPS, respectively. Our primary focus in this study is to underscore the potential of active RIS-assisted MIMO-OFDM systems through practical experimentation, emphasizing the pivotal role of active RIS in MIMO systems. Consequently, we did not implement the power adjustment function. For our hardware components, the magnitude of $\Phi$, without considering the loss of the connection line, provides approximately an 8 dB gain due to a 9 dB loss from the PS and a 3 dB loss from the DPS, and the phase of $\Phi$ can be controlled by the DPS.

\subsection{System Model}
Next, we present the system model with the active RIS, aiming to introduce the corresponding performance metric of interest. We consider a 5G NR-compatible OFDM system.
We model the channel on a per-subcarrier basis, considering the number of antennas at the transmitter (Tx), denoted as $N_{\rm t}$, and at the receiver (Rx), denoted as $N_{\rm r}$. In our designed active RIS, we use $K_{\rm ris}$ active RIS elements, resulting in $K = K_{\rm ris} \times N_{\rm d}$ antennas for the tRIS, as discussed in Section~II.A.
Considering these factors, we can calculate the total power consumption of the RIS as follows:
\begin{equation}
P = K_{\rm ris} \times P_{\rm lna} + K \times P_{\rm dps},
\end{equation}
where $P_{\rm lna}$ and $P_{\rm dps}$ represent the static power consumed by each LNA and DPS, respectively.

The transmit signal from the $n_{\rm t}$-th Tx antenna at one of the subcarriers is represented as $s^{n_{\rm t}}$. The transmitted signal vector can be denoted as $\mathbf{s} = [s^{1}, \ldots, s^{N_{\rm t}}]^T$. Following the approach in \cite{Chian-2023}, the channel frequency response (CFR) is modeled by considering three types of paths in the RIS scenario:
\begin{itemize}
\item First, $s^{n_{\rm t}}$ propagates through the line-of-sight (LoS) path or paths that do not interact with the RIS.
\item Second, when $s^{n_{\rm t}}$ is received by the $k_{\rm r}$-th rRIS and retransmitted by the $k_{\rm t}$-th tRIS, it undergoes propagation via the antenna mode (AM) of the RIS path.
\item Third, when $s^{n_{\rm t}}$ impinges on the metal plate of the $k_{\rm r}$-th rRIS or $k_{\rm t}$-th tRIS and generates reflected and scattered signals towards the receiver, it experiences propagation via the structural mode (SM) of the RIS path.
\end{itemize}
The channel model for the second and third types of paths is illustrated in Fig. \ref{fig:RISelement}(b).

Consequently, the CFR of all paths at one of the subcarriers for the MIMO system is formulate as $\qH = \qH_{\rm los} + \qH_{\rm am} + \qH_{\rm sm}$,
where $\qH_{\rm los}$, $\qH_{\rm am}$, and $\qH_{\rm sm}$ correspond to the three types of paths, respectively. To specify the AM of the RIS path, we use $\qH^{\sf r}_{\rm am} \in \bbC^{N_{\rm r} \times K}$ and $\qH^{\sf t}_{\rm am} \in \bbC^{K_{\rm ris} \times N_{\rm t}}$ to denote the AM of the RIS paths corresponding to the Rx and Tx, respectively. Then, the AM of the RIS path can be expressed as:
\begin{equation} \label{eq:CFR_AM}
    \qH_{\rm am} = {{\qH^{\sf r}_{\rm am}} } \cdot \qPhi \cdot {\left( {\bf 1}_{ N_{\rm d}} \otimes \qH^{\sf t}_{\rm am} \right) },
\end{equation}
where ${\qPhi = {\rm diag}(\Phi^1, \ldots, \Phi^K)}$ with $\Phi^k$ representing the total transmission coefficient of the $k$-th tRIS element as described in \eqref{eq:TC}, and ${\bf 1}_{ N_{\rm d}}$ represents an all-ones vector with $N_{\rm d}$ dimensions. If ${N_{\rm d} = 1}$,  \eqref{eq:CFR_AM} simplifies to the common case of one rRIS paired with one tRIS. Given the CFR and the signal paths, the received signal at the MIMO receiver is given by \cite{Zhang-23TCOM}:
\begin{equation} \label{eq:MIMO}
    \qy = \qH \qs + {\qH^{\sf r}_{\rm am}} \qPhi \qv + \qz,
\end{equation}
where $\qv \sim \mathcal{CN}({\bf 0}_K, \sigma_{\rm v}^2 \qI_K)$ and $\qz \sim \mathcal{CN}({\bf 0}_{N_{\rm r}}, \sigma_{\rm z}^2 \qI_{N_{\rm r}})$ represent the noise introduced by the active RIS and the Rx, respectively.

Since $\qv$ undergoes transmission through the MIMO channel, $\qn_{\rm d} \triangleq {\qH^{\sf r}_{\rm am}} \qPhi \qv$ represents the colored noise. The total noise covariance matrix, denoted as  $\qn \triangleq \qn_{\rm d} + \qz$, can be written as $\qR_{\rm n} = \bbE \{\qn \qn^H\} = \qR_{\rm v} + \sigma_{\rm z}^2 \qI_{N_{\rm r}}$, where $\qR_{\rm v}$ represents the covariance matrix of $\qn_{\rm d}$. To simplify system performance derivations, we multiply $\qy$ by $\qR_{\rm n}^{- {1}/{2}}$ to whiten $\qn$, resulting in
\begin{equation} \label{eq:whiteMIMO}
    \widetilde{\qy} =
    \qR_{\rm n}^{- {1}/{2}} \qH
    \qs +
    \qR_{\rm n}^{- {1}/{2}} \qn
    = \widetilde{\qH} \qs + \widetilde{\qn},
\end{equation}
with $\widetilde{\qH} \triangleq \qR_{\rm n}^{- {1}/{2}} \qH $ , $\widetilde{\qn} \triangleq \qR_{\rm n}^{- {1}/{2}} \qn$, and $\widetilde{\qn} \sim \mathcal{CN}({\bf 0}_{N_{\rm r}}, \qI_{N_{\rm r}})$.

Assuming that the transmission power for each Tx antenna is $1$, the MIMO channel capacity and the SNR at one of the subcarriers after whitening the noise are defined as follows:
\begin{subequations} \label{eq:MIMOpar}
\begin{align}
    {\sf C}
    &=
    \log_2 \det {\left( \qI_{N_{\rm r}} + \widetilde{\qH} \widetilde{\qH}^H
    \right)},
    \label{eq:MIMOpar_Cap}                                                       \\
    {\sf SNR}
    &=
    \tr
    {( \widetilde{\qH}  \widetilde{\qH}^H
    )} / N_{\rm r}.
    \label{eq:MIMOpar_SNR}
\end{align}
\end{subequations}
In addition to capacity and SNR, the bit error rate (BER) is also of interest. To evaluate this metric, we employ the expectation propagation-based (EP) MIMO detector \cite{Zhang-22JSAC} for both simulations and experiments.

For the simulations, we generate the elements of all the channel matrices, including $\qH_{\rm los}$, $\qH_{\rm sm}$, $\qH^{\sf r}_{\rm am}$, and $\qH^{\sf r}_{\rm am}$, based on the 3GPP spatial channel model (SCM) with antenna and polarization effects \cite{Chian-2023}.
For the experiments, we have developed a prototype that is described in detail in the following subsection. This prototype allows us to validate the mentioned metrics through real-world scenarios.

\subsection{Active RIS-Assisted MIMO System}
\begin{figure}
    \centering
    \resizebox{3.5in}{!}{%
    \includegraphics*{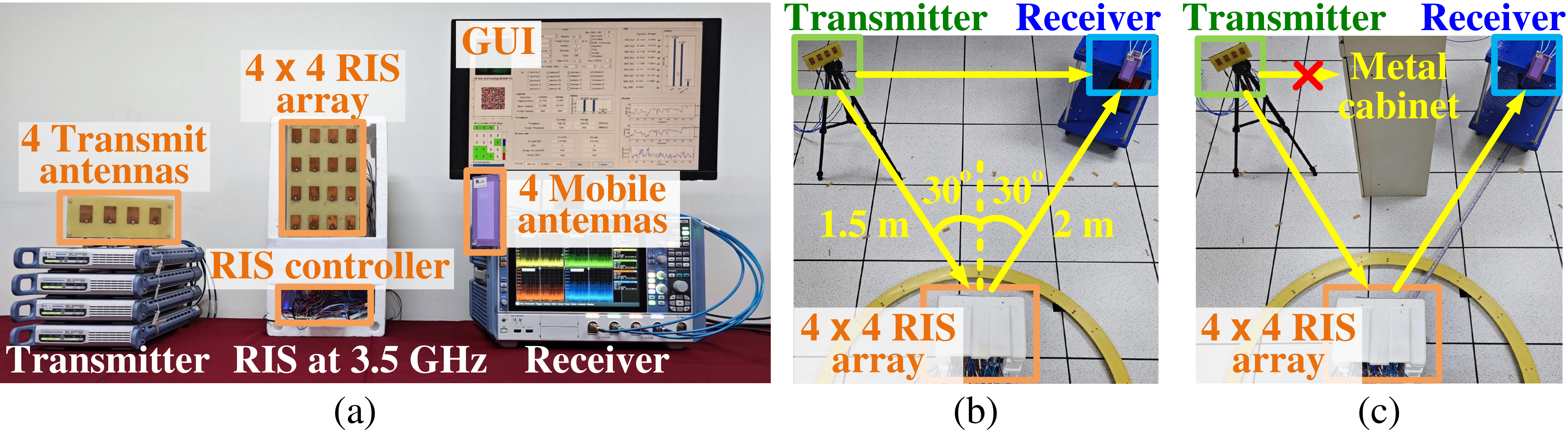} }%
    \caption{(a) Prototype of the 4T4R MIMO-OFDM system with assistance from a $4 \times 4$ active RIS array. Experimental setups: (b) with LoS and (c) without LoS.
    \label{fig:Prototype_MeasScenario}}
    \vspace{-0.5cm}
\end{figure}

In this subsection, we present the prototype of our active RIS-assisted 4T4R MIMO-OFDM system, as depicted in Fig.~\ref{fig:Prototype_MeasScenario}(a), which can be categorized into three components: the transmitter (Tx), RIS, and receiver (Rx).

For the Tx, we transmit four-stream 5G OFDM signals in the 3.45-3.55 GHz band over the air.
For our setup, we employ four one-port R\&S vector signal generators, specifically the {\tt SGT100A} model, which are connected to four transmit patch antennas. These antennas are positioned at intervals of half the wavelength corresponding to the 3.5 GHz frequency.

The RIS employs a $4 \times 4$ active RIS array, with an antenna structure similar to \cite{Chian-2023}. The detailed structure of each active RIS element is described in Section II-A. We use a DE10-Nano Kit, based on a Cyclone V SoC FPGA, as the RIS controller.

For the Rx, we receive the 5G signals using four mobile antennas, which are downconverted to the baseband using a four-port R\&S digital oscilloscope, specifically the {\tt RTP164}. The baseband signals are processed using a C program on a personal computer. Communication between the computer and the RIS controller takes place via a half-duplex Bluetooth connection, enabling the determination of the controlling strategy.

The framework of our MIMO-OFDM system in the baseband is as follows. The encoded data is converted to four-stream 5G signals using OFDM modulation. We utilize a bandwidth of $100$ MHz with a subcarrier spacing of $60$ kHz. The signals are subsequently upconverted to the $3.5$ GHz band and transmitted. The receiver downconverts the received signals to the baseband, and synchronization is performed using 5G synchronization signal blocks. By performing channel estimation, we can calculate the capacity \eqref{eq:MIMOpar_Cap} and SNR \eqref{eq:MIMOpar_SNR} of the 4T4R MIMO-OFDM system. After performing the EP MIMO detector, we obtain the uncoded BER. Note that our system adheres to the 5G NR standards, so the experimental performance and properties of the active RIS-assisted MIMO-OFDM system can be referenced in the context of 5G communications.

\section*{III. Proof of Concept and Performance Evaluation}

\subsection*{A. Controlling Algorithms}

To control the RIS, we employ the BG algorithm \cite{Chian-2023}, which consists of two main steps: Random-Max Sampling (RMS) and Greedy Searching (GS). In RMS, a few codewords are drawn uniformly from an exhaustive codebook, and the best codeword is selected as the initial stage for the subsequent GS step. In GS, a progressive search is performed from the exhaustive states of each phase shifter, iterating through all the RIS elements. Instead of selecting the best received signal quality as in \cite{Chian-2023}, we choose the RIS's state corresponding to the highest channel capacity in \eqref{eq:MIMOpar_Cap} using the greedy method. The complexity of BG is represented as $O(KM)$, where $K$ represents the number of DPS, and $M$ is the number of phase selections available for each DPS.

In addition to BG, we consider two other controlling algorithms for comparison. The first algorithm is CSM \cite{Ren-2022}, which involves drawing uniformly from an exhaustive codebook and determining the RIS's state based on the phase expectation of each RIS element. We set the number of CSM's random samples to be 8 times the number of RIS elements.

The second algorithm is the conventional codebook-based algorithm, named the maximum-power codebook (MPC), which relies on the determined beam from the RIS. In MPC, we replace the Tx and Rx antennas in Fig. \ref{fig:Prototype_MeasScenario}(b) with single horn antennas. With the RIS located at the center of the half circle with a radius of 2\,m, we determine the RIS's codebooks by applying BG with the SNR in the SISO system to the Rx antenna locations along the half circle spaced by $5^{\circ}$. The determined codebooks, corresponding to different directions of RIS beams, are then utilized by MPC to select the best capacity among them in the MIMO system.

Note that the best RIS codeword in all the mentioned algorithms is determined by probing the capacity performance based on a set of codewords, and they do not require explicit channel matrix information. The performance of these algorithms depends on the number of probes, where a higher number of probes generally yields better performance. Therefore, their performances should be compared under the same number of probes.

\subsection*{B. Experiments}

\begin{figure} [t]
    \centering
    \includegraphics[width=3.5in]{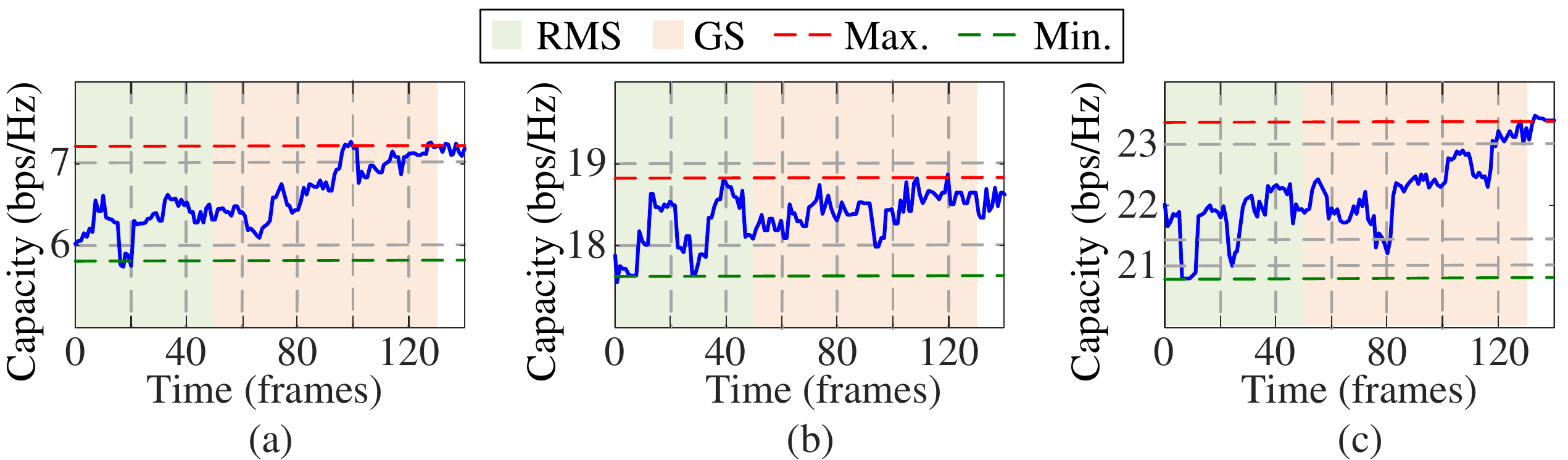}
    \caption{State evolution of the controlling process using BG for: (a) passive RIS in a 1T4R OFDM system, (b) passive RIS in a 4T4R MIMO-OFDM system, and (c) active RIS in a 4T4R MIMO-OFDM system.}
    \label{BG}
    \vspace{-0.5cm}
\end{figure}

\begin{table*}
\begin{center}
\begin{footnotesize}
\begin{threeparttable}
\caption{Comparison of 4T4R MIMO system in the different cases. \label{tab:ComparMIMO}}
\begin{tabular}{|l||c|ccc|ccc|cc|c|ccc|}
\hline

\multirow{3}{*}{Case} &
\multicolumn{9}{|c|}{(64 QAM, w/o LoS)} &
\multicolumn{4}{|c|}{(256 QAM, with LoS)} \\ \cline{2-14}

    & w/o &
\multicolumn{3}{|c}{(Passive, BG)}  &
\multicolumn{3}{|c}{(Active, BG)}   &
\multicolumn{2}{|c|}{(Active, MPC)}
    & w/o &
\multicolumn{3}{|c|}{(Active, BG)}   \\
    & RIS &
Initial & Best  & Diff. &
Initial & Best  & Diff. &
Best    & Diff.
    & RIS &
Initial & Best  & Diff.  \\ \hline \hline

SNR (dB) &
\multirow{1}{*}{18.23} &
\multirow{1}{*}{18.55} & \multirow{1}{*}{18.88} & \multirow{1}{*}{0.45} &
\multirow{1}{*}{21.40} & \multirow{1}{*}{22.93} & \multirow{1}{*}{1.61} &
\multirow{1}{*}{22.01} & \multirow{1}{*}{1.29} &
\multirow{1}{*}{24.40} &
\multirow{1}{*}{25.10} & \multirow{1}{*}{25.26} & \multirow{1}{*}{0.21} \\
\hline

Capacity (bps/Hz)&
\multirow{1}{*}{17.82} &
\multirow{1}{*}{18.37} & \multirow{1}{*}{18.91} & \multirow{1}{*}{0.60} &
\multirow{1}{*}{21.98} & \multirow{1}{*}{22.68} & \multirow{1}{*}{1.27} &
\multirow{1}{*}{22.10} & \multirow{1}{*}{0.91} &
\multirow{1}{*}{24.36} &
\multirow{1}{*}{28.74} & \multirow{1}{*}{30.05} & \multirow{1}{*}{1.59} \\
 \hline

Uncoded BER&
\multirow{1}{*}{0.093} &
\multirow{1}{*}{0.088} & \multirow{1}{*}{0.074} & \multirow{1}{*}{0.016} &
\multirow{1}{*}{0.048} & \multirow{1}{*}{0.029} & \multirow{1}{*}{0.032} &
\multirow{1}{*}{0.036} & \multirow{1}{*}{0.039} &
\multirow{1}{*}{0.095} &
\multirow{1}{*}{0.046} & \multirow{1}{*}{0.024} & \multirow{1}{*}{0.029} \\
 \hline

SE (bps/Hz)&
\multirow{1}{*}{- \tnote{$\dagger$}} &
\multirow{1}{*}{-} & \multirow{1}{*}{-} & \multirow{1}{*}{-} &
\multirow{1}{*}{18.76} & \multirow{1}{*}{19.13} & \multirow{1}{*}{-} & \multirow{1}{*}{18.97} & \multirow{1}{*}{-} &
\multirow{1}{*}{-} &
\multirow{1}{*}{25.05} & \multirow{1}{*}{25.64} & \multirow{1}{*}{-} \\
       \hline

\end{tabular}
\begin{tablenotes}
    \item[$\dagger$]
    If an uncoded BER is greater than 5e-2, SE will not be considered.
\end{tablenotes}
\end{threeparttable}
\end{footnotesize}
\end{center}
\vspace{-0.5cm}
\end{table*}

We performed experimental measurements in an open area of the NSYSU Electrical and Computer (EC) Eng. Building to assess our active RIS-assisted MIMO-OFDM system. The experimental setup is illustrated in Figs. \ref{fig:Prototype_MeasScenario}(b) and \ref{fig:Prototype_MeasScenario}(c). In this setup, the RIS was positioned 1.5 m away from the Rx and 2 m from the Tx. We examined two scenarios: the first where the LoS path was obstructed by a metal cabinet, as shown in Fig.~\ref{fig:Prototype_MeasScenario}(c), and the second where the LoS path remained unblocked, as depicted in Fig.~\ref{fig:Prototype_MeasScenario}(b).

To analyze the difference between passive RIS and active RIS in the BG optimization process, we present the state evolution of the controlling process in both the SIMO (1T4R) and MIMO (4T4R) systems in Fig. \ref{BG}. Two levels of phase shift  were chosen. The structure of the passive RIS was the same as described in \cite{Chian-2023}. In the SIMO system with the passive RIS, the capacity, as depicted in Fig. \ref{BG}(a), undergoes significant changes during RMS and shows a smooth increase during GS. However, in the MIMO system with the passive RIS, the capacity gain during GS, as shown in Fig. \ref{BG}(b), is minimal. In contrast, when the passive RIS is replaced by the active RIS in the same scenario, the capacity gain during GS, as shown in Fig. \ref{BG}(c), becomes more pronounced due to the use of amplifiers.

Furthermore, we compare the overall performance of the passive RIS and active RIS with and without the LoS path in Table \ref{tab:ComparMIMO}.The SNR indicates the quality of the received signal. In the contexts of SISO and SIMO, a higher SNR typically translates to a greater system throughput or spectral efficiency. However, in MIMO scenarios, SNR is not the sole determinant of system throughput. The conditions of the channel rank play a significant role as well. This leads us to consider channel capacity, which represents the maximum theoretical transmission data rate. To get a more comprehensive view of system performance, we also examine the BER which offers a direct measure of the system throughput for the entire design, encompassing all associated impairment losses. Our target is to achieve an uncoded BER of 5e-2, as a decoded BER using LDPC coding with a 1/2 code rate can approach 0 with this uncoded BER. To analyze the optimization process, we consider three types of RIS states, shown in Table \ref{tab:ComparMIMO}. The first type is the initial state before applying BG. The second type is the final state after BG, which corresponds to the best state of the RIS. The third type represents the difference between the best and worst states of the RIS encountered during BG, which reflects the RIS's control ability. The performance of the initial RIS state falls between the performance of the best and worst RIS states.

In all cases without RIS, communication is not supported as the uncoded BER exceeds 5e-2. In the case without LoS, the passive RIS is not suitable for the MIMO-OFDM system as the performance gain is weak. However, even with the active RIS in the initial state, the performance gain is significant, and the BER can be reduced below 5e-2. Moreover, the improved controlling ability allows the MIMO-OFDM system to be more robust in different environmental conditions. To analyze the effect of multipath, we compare BG with MPC, as MPC is similar to using RIS with beamforming techniques. The performance of MPC is worse than that of BG, and the final codebook of MPC corresponds to a different beam expectation, specifically $45^{\circ}$ instead of $30^{\circ}$ in Fig. \ref{fig:Prototype_MeasScenario}. This result indicates that considering the role of the RIS as a single-path assistance in MIMO scenario may not be appropriate. Finally, the controlling ability of SNR in the case with LoS (0.21\,dB Diff.) is noticeably worse than that in the case without LoS (1.61\,dB Diff.), as the SNR without RIS in the LoS case is much better than without RIS in the non-LoS case. However, the active RIS maintains significant control over capacity, demonstrating its capability to enhance the multipath channel.

\subsection*{C. Simulations}

\begin{figure} [t]
    \centering
    \includegraphics[width=3.5in]{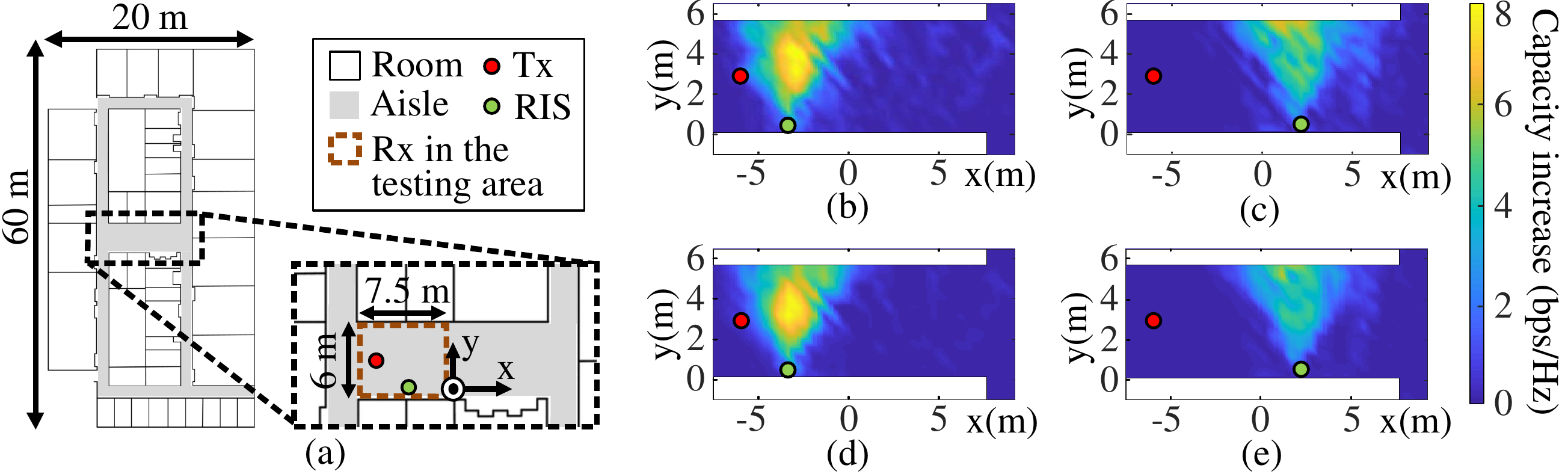}
    \caption{(a) The simulation environment. Capacity increase using (b), (c) $2 \times 2$ active RIS, and (d), (e) $16 \times 16$ passive RIS at distinct locations}
    \label{SimScenario}
    \vspace{-0.5cm}
\end{figure}

To further analyze the properties of the RIS-assisted MIMO system, simulations are conducted using Wireless Insite$^\circledR$ software, which incorporates the obtained channel impulse response into the channel model. The simulation environment, shown in Fig.~\ref{SimScenario}(a), represents a scenario similar to the experiment described in Section III.B. The system also corresponds to the 5G NR 4T8R MIMO-OFDM system, with Tx antennas oriented along the $x$-axis and RIS antennas facing along the $y$-axis. Four levels of phase shift were considered.

First, to assess the impact of active RIS placement on MIMO system performance, we conducted simulations with a $2 \times 2$ active RIS with 18 dB power gain in Figs. \ref{SimScenario}(b) and \ref{SimScenario}(c) and $16\times 16$ passive RIS in Figs. \ref{SimScenario}(d) and \ref{SimScenario}(e), positioned at two distinct coordinates. We observed the most substantial system performance enhancement when the RIS was situated as in Figs. \ref{SimScenario}(b) and (d). Positions shown in Figs.~\ref{SimScenario}(c) and~(e) yielded the least optimal results. This suggests that the performance gains offered by both active and passive RIS are influenced by their distance from the BS. Nearer placements prove to be more beneficial.\footnote{For optimal wireless power transfer, \cite{Fu-Mul_ARIS/PRIS} posits a different view, suggesting that the active RIS should be positioned closest to the receiver.}

\begin{figure} [t]
    \centering
    \includegraphics[width=3.5in]{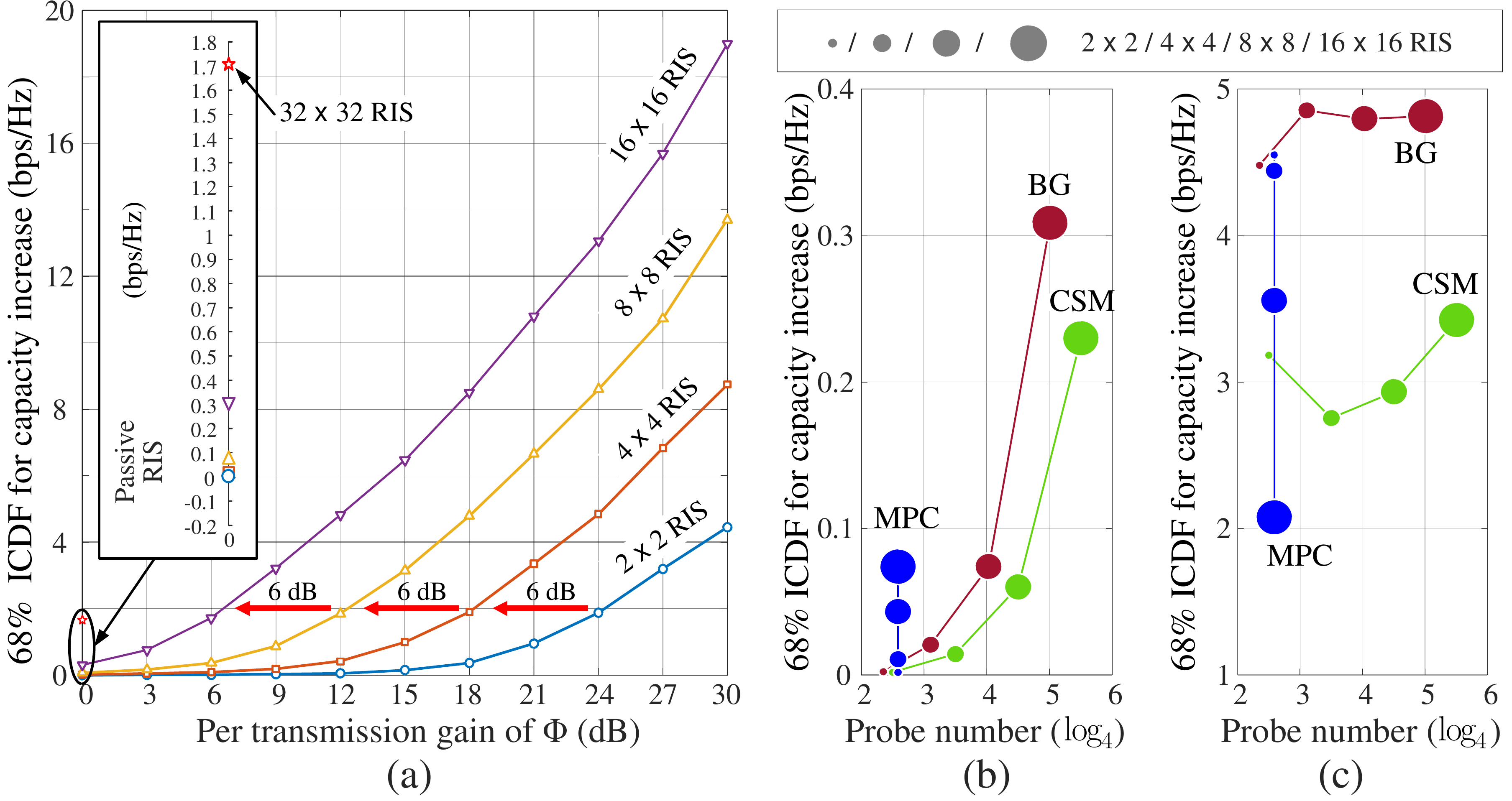}
    \caption{(a) Effects of varying total transmission gains across different sizes of square-arranged RIS arrays. (b) and (c) depict the impact of different controlling algorithms for passive and active RIS, respectively.}
    \label{DiffPower&Algo}
    \vspace{-0.25cm}
\end{figure}

Next, we discuss the performance inferences due to the size of the RIS array and the amplification gain $\Phi$ in \eqref{eq:TC}. We utilized the $68\%$ inverse cumulative distribution function (ICDF) to quantify the capacity improvement brought about by the RIS in the testing area depicted by the red frame in Fig. \ref{SimScenario}(a). In Fig. \ref{DiffPower&Algo}(a), the capacity increase achieved by BG is compared when using variable amplification gain $\Phi^k$ to each RIS element. Interestingly, by doubling the dimensions of the RIS array, we achieve performance akin to adding a 6 dB transmission gain, attributed to the increased number of RIS elements. Specifically, to achieve an equivalent capacity increase to that of a $16 \times 16$ passive RIS, the $2 \times 2$, $4 \times 4$, and $8 \times 8$ active RISs require power gains of 18 dB, 12 dB, and 6 dB, respectively. These gains correspond to power consumptions of 189 mW, 47.5 mW, and 12 mW, which all larger than the 1.28 mW of the $16 \times 16$ passive RIS. However, these active RIS configurations necessitate only 1/64, 1/16, and 1/4 of the complexity inherent in a $16 \times 16$ passive RIS. Such a decrease in complexity impacts the number of probes, thereby influencing the optimization time. Consequently, while active RIS configurations offer faster optimization times, they are less energy efficient than passive RIS. It appears a trade-off between the energy efficiency and the optimization time.

To further analyze the optimization efficiency for different controlling algorithms, we consider their probe numbers and capacity increase. The results for passive and active RIS arrays of different sizes are shown in Figs. \ref{DiffPower&Algo}(b) and \ref{DiffPower&Algo}(c), respectively. The probe number is represented using a base-4 logarithm, and the maximum number of probes depends on the number of RIS elements. In the figure, a larger value on the x-axis indicates a longer searching time, while a larger value on the y-axis represents higher algorithm performance. Therefore, the most efficient algorithm should ideally appear in the top-left corner, where it achieves high performance with a short search time.

For the passive RIS, as the RIS array size increases, the capacity growth for MPC is notably smaller than for other algorithms, even though its number of probes remains constant. Additionally, CSM requires a larger number of random probes to achieve good performance, resulting in higher complexity compared to the other algorithms. Indeed, none of the algorithms reside in the top-left corner, indicating inefficiency with passive RIS.

Regarding the active RIS, we impose a restriction on the sum of the total transmission gains, limiting it to 36\,dB. As such, the amplification gains for a ${2 \times 2}$ RIS element and a ${16 \times 16}$ RIS element are 30 dB and 12 dB, respectively. As the number of RIS elements increases, the capacity increase achieved by MPC diminishes. This outcome suggests that relying solely on the conventional beamforming-based codebook, which focuses only on optimizing a single path, does not adequately support the MIMO system. These findings align with the experimental results presented in Table \ref{tab:ComparMIMO}. Moreover, small active RISs, which belong to the top-left corner of the performance plot, prove to be an efficient choice in the context of the MIMO system.

\section*{IV. Conclusion}

This study presented the first experimental development of an active RIS-assisted MIMO-OFDM system in the 5G NR context, utilizing practical BS and smartphone form factors. Our results underscored the pivotal role of active RIS in enhancing MIMO systems. Nevertheless, a trade-off between energy efficiency and optimization time became evident. We also demonstrated that the conventional beamforming-based codebook falls short in supporting the MIMO system effectively. Among the algorithms evaluated, BG exhibited superior optimization efficiency, especially in terms of probe numbers and capacity increase. The quest for a more intelligent controlling algorithm for MIMO systems remains a critical challenge.



\end{document}